\documentclass{webofc}
\usepackage[varg]{txfonts}   
\usepackage{hyperref}
\begin{document}
\title{Implementation of New Security Features in CMSWEB Kubernetes Cluster at CERN}
%
%

\author{\firstname{Aamir} \lastname{Ali}\inst{1}\fnsep\thanks{\email{aamir.ali@cern.ch}} 
\and
        \firstname{Muhammad} \lastname{Imran}\inst{2}\fnsep\thanks{\email{muhammad.imran@ncp.edu.pk}} \and
        \firstname{Valentin} \lastname{Kuznetsov}\inst{3}\fnsep\thanks{\email{vkuznet@protonmail.com}}
             \and
        \firstname{Spyridon} \lastname{Trigazis}\inst{1}\fnsep\thanks{\email{Spyridon.Trigazis@cern.ch}}
             \and
        \firstname{Aroosha} \lastname{Pervaiz}\inst{2}\fnsep\thanks{\email{aroosha.pervaiz@cern.ch}}
                  \and
        \firstname{Andreas} \lastname{Pfeiffer}\inst{1}\fnsep\thanks{\email{andreas.pfeiffer@cern.ch}}
             \and
                     \firstname{Marco} \lastname{Mascheroni} 
\inst{4}\fnsep for CMS Collaboration \thanks{\email{marco.mascheroni@cern.ch}}
}
\institute{CERN, Geneva, Switzerland. 
\and
           National Centre for Physics, Islamabad, Pakistan.
\and
           Cornell University, USA. 
\and
           University of California San Diego, USA.            
          }

\abstract{%
The CMSWEB cluster is pivotal to the activities of the Compact Muon Solenoid (CMS) experiment, as it hosts critical services required for the operational needs of the CMS experiment. The security of these services and the corresponding data is crucial to CMS. Any malicious attack can compromise the availability of our services. Therefore, it is important to construct a robust security infrastructure. In this work, we discuss new security features introduced to the CMSWEB Kubernetes (“k8s”) cluster. The new features include the implementation of network policies, deployment of Open Policy Agent (OPA), enforcement of OPA policies, and the integration of Vault. The network policies act as an inside-the-cluster firewall to limit the network communication between the pods to the minimum necessary, and its dynamic nature allows us to work with microservices. The OPA validates the objects against some custom-defined policies during create, update, and delete operations to further enhance security. Without recompiling or changing the configuration of the Kubernetes API server, it can apply customized policies on Kubernetes objects and their audit functionality enabling us to detect pre-existing conflicts and issues. Although Kubernetes incorporates the concepts of secrets, they are only base64 encoded and are not dynamically configured. This is where Vault comes into play: Vault dynamically secures, stores, and tightly controls access to sensitive data. This way, the secret information is encrypted, secured, and centralized, making it more scalable and easier to manage. Thus, the implementation of these three security features corroborate the enhanced security and reliability of the CMSWEB Kubernetes infrastructure.}
\maketitle
%
\section{Introduction\label{intro_sect}}

The CMS experiment\cite{collaboration2008cms,cms2023development} simulates, reconstructs, and analyzes the data collected during collision runs by running hundreds of thousands of jobs on its distributed computing system. Essential CMS central services that handle CMS data administration, data discovery, and various data bookkeeping activities are hosted on a separate cluster called CMSWEB. To reduce the release upgrade cycle and complete end-to-end deployment procedures for CMSWEB services, CMS migrated its infrastructure \cite{imran2021migration} to a containerized solution based on Docker \cite{rad2017introduction} orchestrated with Kubernetes \cite{luksa2017kubernetes}. With this approach, CMSWEB not only unified deployment procedures and reduced the release upgrade cycle but also reduced operational costs significantly. 

Security in a Kubernetes cluster is of paramount importance as it safeguards against data breaches, unauthorized access, and potential downtime. Kubernetes manages complex containerized applications, making it a high-value target for attackers. A secure Kubernetes cluster ensures the integrity, confidentiality, and availability of applications and data. It involves strategies like role-based access control (RBAC), network policies, container image scanning, and secrets management. Neglecting security can lead to serious consequences, including data leaks or loss, service disruptions, and reputational damage. Therefore, prioritizing security in a Kubernetes cluster is essential to maintain trust, compliance, and the overall health of your containerized infrastructure.

To enhance the security of the CMSWEB k8s cluster, we incorporated some new security features. The first feature is the implementation of \textit{network policies}. The network policies in Kubernetes are a set of rules that control the communication between pods within a cluster. They define how pods can communicate with each other and with external resources based on IP addresses, ports, and labels. Network policies help enhance security by specifying which network traffic is allowed or denied, allowing for fine-grained control over pod-to-pod communication. The second feature is incorporation of \textit{Open Policy Agent (OPA) Gatekeeper policies}. This feature allows us to enforce policies and constraints on Kubernetes resources, such as pods and deployments, by defining custom policies using Rego, a policy language. These policies can be used to ensure compliance, security, and best practices within Kubernetes cluster, providing a way to validate and control the configuration and behavior of resources. The third feature is the integration of \textit{Vault} in our cluster. Vault is crucial for ensuring the security of sensitive data, such as API keys, passwords, and encryption keys, in modern IT infrastructures. It provides a centralized and secure way to store, access, and manage these secrets, enforcing access control, auditing, and rotation policies. 

The rest of this paper is organized as follows: Section \ref{chap:networkpolicies} gives an overview of network policies and describes how these policies have been implemented. Section \ref{chap:opa} introduces Open Policy Agent and describes how it is used to implement fine-grained control over user requests. Section \ref{chap:vault} gives a brief introduction to Vault and how it has been used to provide secret management service for CMSWEB cluster. Finally, we conclude in Section \ref{chap:Conclusion}. 

\section{Network Policies}
\label{chap:networkpolicies}
In this section, we give a brief overview of network policies and how we implemented these policies in the context of CMSWEB k8s clusters. Network traffic is governed by a set of rules known as network policy. A network policy \cite{kubernetes_np} in Kubernetes enables the administrator to manage traffic flow at the level of the IP address, port and labels. It is an application-centric construct that enables determining the range of network entities with which a pod may communicate. Combining the three identifiers—other pods that are allowed, namespaces that are allowed, and IP blocks that are allowed—identifies the entities that a pod can connect with. It is applied to a connection with a pod to either incoming or outgoing or both types of connections and is not relevant to other connections. By default, Kubernetes does not restrict traffic between pods running inside the cluster. So, if one pod is compromised, all pods are potentially compromised. Therefore, a network policy must be in place to avoid such circumstances. 

\subsection{Implementation}
The CMSWEB cluster uses \textit{Calico} as a network plugin to k8s that supports network policy enforcement. There are two types of network policies based on the type of traffic they control; \textit{Ingress} controls the incoming traffic while \textit{Egress} controls the outgoing traffic \cite{kubernetes_np}. 

CMSWEB currently has 3 different environments, \textit{test}, \textit{pre-production}, and \textit{production environment}. As far as the backend is concerned, all the services are the same for all 3 environments but there are some minor differences in the frontend of these environments. In the test environment, there are 3 services; \textit{auth-proxy-server}, \textit{scitokens-proxy-server}, and \textit{x509-proxy-server}. In the pre-production and production environment, there is only one service; \textit{frontend}, and \textit{nginx-ingress} respectively. 

\begin{figure*}[!t]
\includegraphics[scale=0.55]{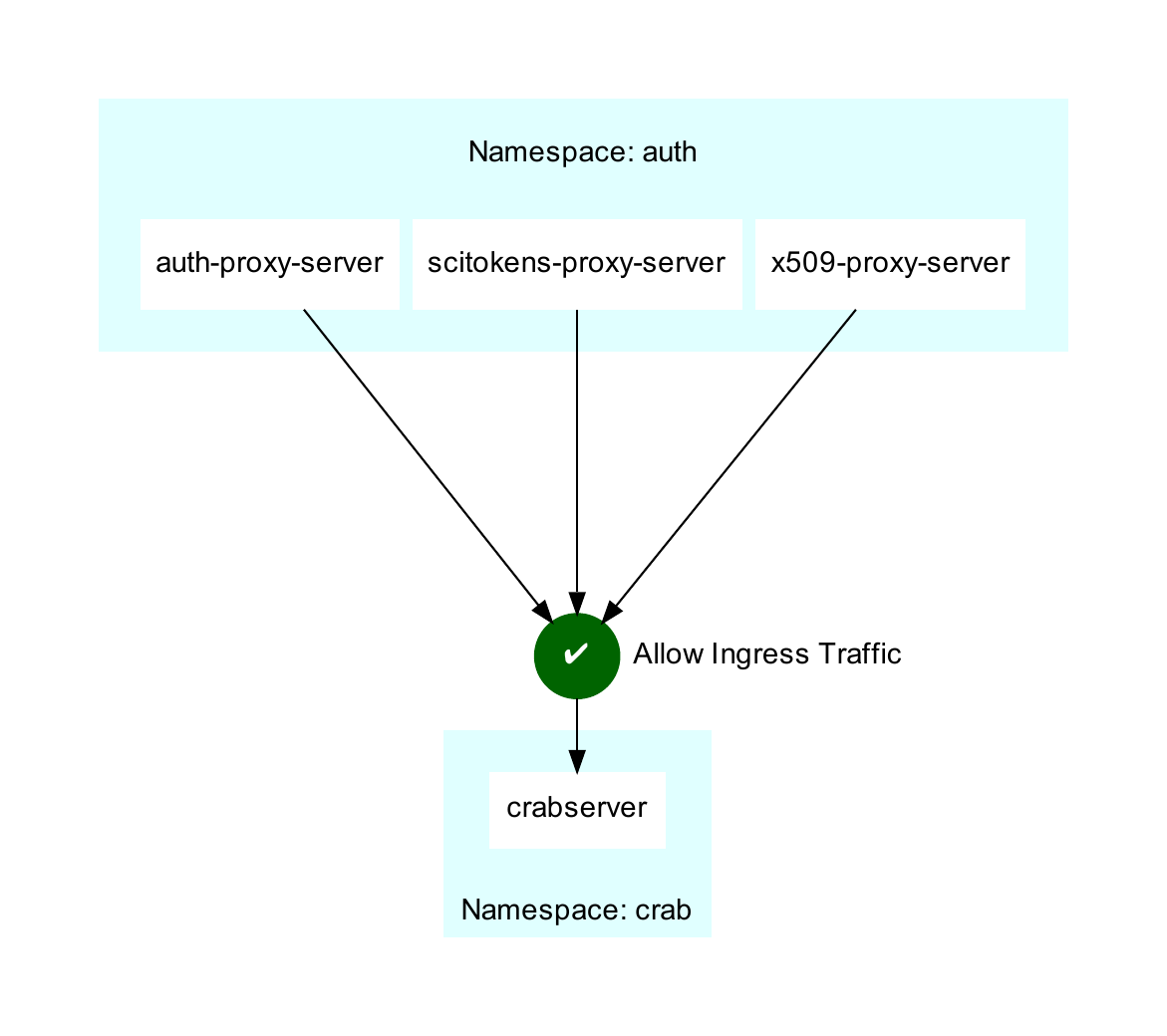}
\centering
\caption{Network policy of crab namespace in CMSWEB Test environment}
\label{fig:crab}
\end{figure*}

Fig \ref{fig:crab} shows an example of a network policy from the test environment. The \textit{crabserver} is a backend service deployed in its own namespace \textit{crab}. It allows ingress traffic from all 3 services in \textit{auth} namespace. All the traffic from other namespaces and even other services from auth namespace is blocked, as is outgoing traffic. Similar policies were implemented in the pre-production and production environment with the frontend service. In short, backend services will only accept traffic from frontend services etc.

\section{Open Policy Agent (OPA)}
\label{chap:opa}
This section gives a brief insight on OPA gatekeeper policies and how we incorporated these polices in CMSWEB k8s cluster. OPA \cite{opa} is an open-source general-purpose policy engine which allows consistent, context-aware policy enforcement throughout the entire stack. It separates the formulation of policy from its implementation. OPA offers a high-level declarative language ("Rego") that makes it possible to describe the policy as code as well as straightforward APIs to remove the burden of making policy decisions from the software. The software asks OPA and provides structured data (for example, JSON) as input when making choices about policies. OPA accepts input of any structured data.

In the Kubernetes cluster, when we perform any create, update, and delete operations on objects, the \textit{admission controllers} \cite{kubernetes_admission} enforce policies. Through the use of admission controller webhooks, Kubernetes enables the decoupling of policy decisions from the inner workings of the API Server. In Kubernetes, the admission controllers are essential for enforcing policies. Incoming objects can also be altered by admission controllers. OPA can be deployed as validating admission controller. Since policy decisions can be any arbitrary structured data, they can also be used as a modifying admission controller. A customized project called OPA Gatekeeper \cite{gatekeeper} offers initial interaction between OPA and Kubernetes. In addition to simple OPA, Gatekeeper also provides Audit functionality, Native Kubernetes custom resource definitions (CRDs) for extending the policy library, an extensible, parameterized policy library, and Native Kubernetes CRDs for instantiating the policy library.

\begin{figure*}[!t]
\includegraphics[scale=0.55]{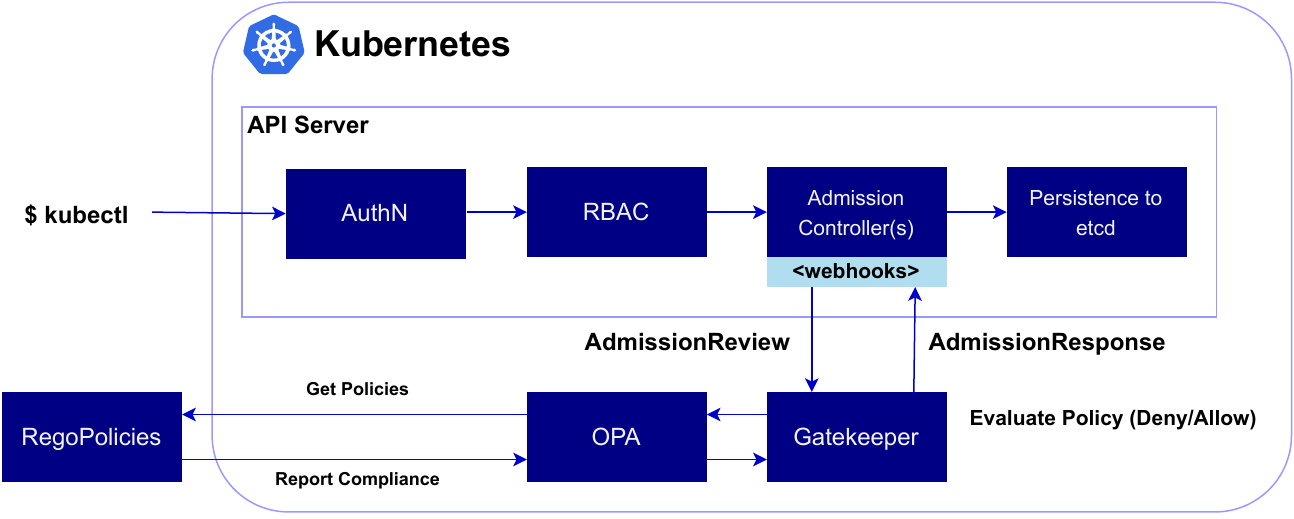}
\centering
\caption{Workflow of Request when OPA is in place}
\label{fig:opa_integration}
\end{figure*}

To avoid the hassle of learning a new language and programming each policy individually, a community-owned collection of Constraint Templates and Constraints is available at \cite{opa_library}. We have used some of the Constraint Templates from this library. A brief description of some policies is given in table \ref{tab:opa_policies};

\begin{table*}[!t]
    \centering
    \begin{tabular}{l|p{8cm}}
\bf{Policy} & \bf{Description} \\  
    \hline  
Allowed Repos & Requires container images to begin with a string from the specified list. \\
Container Limits & Requires that containers have established memory and CPU limits, and that limitation be kept within the designated maximum values. \\
Container Requests & Requires that containers have their memory and CPU requests set, as well as that requests, stay under the permitted upper limits. \\
Container Resource Ratios & Restricts the maximum ratio of container resource limits to requests. \\
Container Resources & Requires containers to have defined resources set. \\
Disallow Anonymous & Prohibits assigning ClusterRole and Role resources to the system:unauthenticated group and system:anonymous user. \\
Replica Limits & Requires that objects (Deployments, ReplicaSets, etc.) with the property "spec.replicas" indicate the number of replicas within specified ranges. \\
Required Probes & Requires the readiness and/or liveness probes to be present in the Pods. \\
Capabilities & Controls Linux capabilities on containers. Corresponds to the 'allowedCapabilities' and 'requiredDropCapabilities' fields in a PodSecurityPolicy \\
Host Namespaces & Prevents pod containers from sharing host PID and IPC namespaces. Corresponds to the parameters "hostPID" and "hostIPC" in a pod security policy. \\
    \hline  

    \end{tabular}
    \caption{Description of OPA Policies}
    \label{tab:opa_policies}
\end{table*}

The Kubernetes package manager \textit{Helm} \cite{helm} makes it simpler to deploy, upgrade, and roll back Kubernetes resources by reducing the procedures to a single CLI command. A collection of files defining k8s resources, a \textit{chart}, is the package format used by Helm. Additionally, Helm enables us to package all of the relevant Kubernetes resources. Thus, packaging, distributing, downloading, and installing these Helm charts is made simple. We created a helm chart for OPA Policies using the constraint templates and related constraints. OPA policies pack all the constraint template library and its constraints as dependencies chart. 

\begin{figure*}[!t]
\includegraphics[scale=0.5]{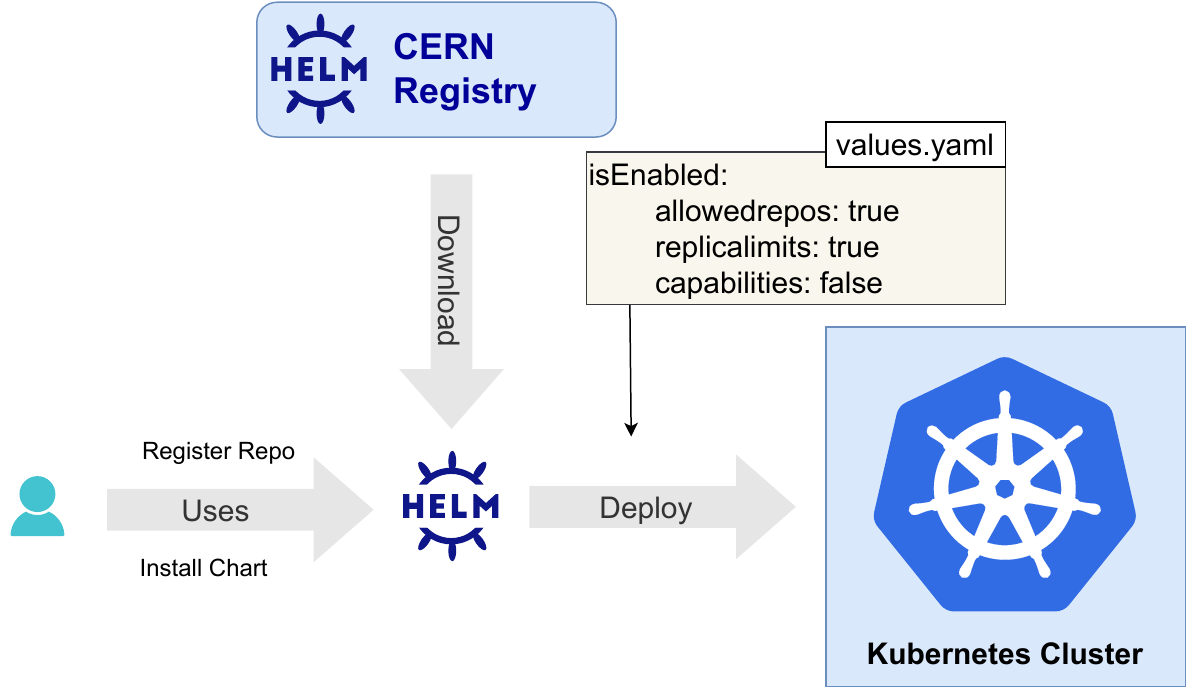}
\centering
\caption{Deployment of OPA policies with Helm}
\label{fig:helm_deploy}
\end{figure*}

Fig \ref{fig:helm_deploy} shows the procedure to deploy the OPA Policies helm chart from the official CERN registry \cite{registry_opapolicies}. Users add the CERN registry as a repository to helm so it can search and install charts from that repository. The user can install the helm chart using helm CLI, helm downloads the chart from the CERN registry, and deploys it to Kubernetes Cluster. During the deployment procedure, helm uses an optional \textit{values.yaml} file if the user passes it as an argument.

The OPA Gatekeeper adds another layer of security. We have configured it to only use the CMSWEB repository to deploy images so nobody can (accidentally) deploy a malicious image. Limiting the number of replicas as well as limiting the resources a pod can use further prevents unfavorable conditions. Requiring readiness and liveness probes makes the admin responsible to ensure the reliability of the service. 

\section{Vault}
\label{chap:vault}

In this section, we discuss the security of k8s secrets and introduce the integration of Vault in the CMSWEB k8s clusters. The \textit{Vault} by HashiCorp \cite{vault} is an identity-based secrets and encryption management system. The Vault tokens, passwords, certificates, and encryption keys are securely stored and access to them is strictly regulated to protect sensitive data. Before granting users, machines, or applications any access to secrets or sensitive data, it verifies and authorizes the clients (users, apps, and machines). Numerous secret engines that either encrypt, store encrypted data, or produce dynamic secrets are supported by Vault. It also supports various authentication engines that provide several methods for logging into Vault. The secret storage and authentication methods added frequently for a variety of situations make Vault extensible.

\begin{figure*}[!t]
\includegraphics[scale=0.6]{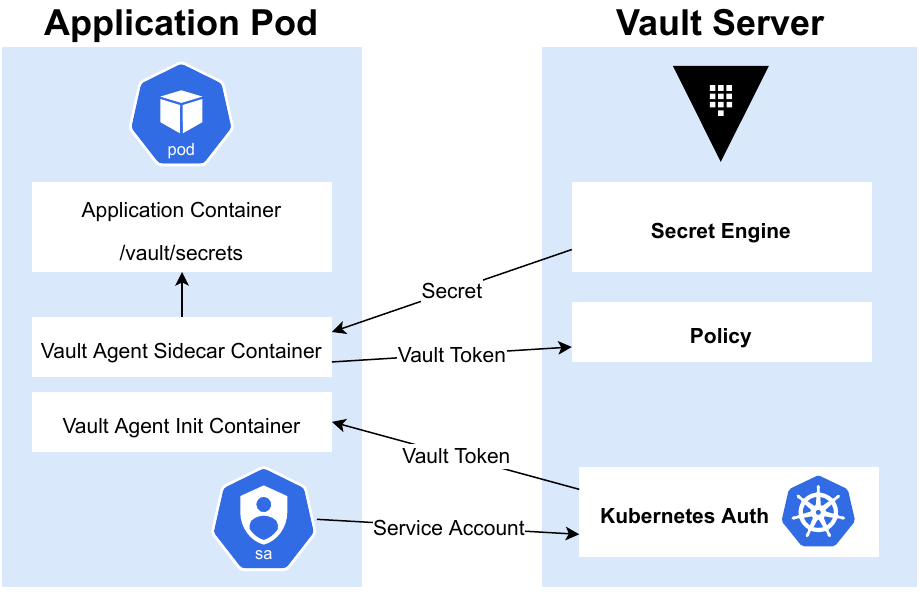}
\centering
\caption{Working of Vault when Integrated with Kubernetes}
\label{fig:vault_k8s_working}
\end{figure*}

In the CMSWEB cluster, we deployed Vault using the helm chart provided by HashiCorp with custom configurations. Vault is deployed as a stand-alone server inside a pod in the vault namespace. The storage is statically created from OpenStack to keep data in case of accidental removal of the Vault server. It’s initialized and unsealed manually, and the Vault root token has been removed from the Vault CLI for security reasons. Kubernetes is then configured as the authenticator. For secrets in CMSWEB, an encryption engine \textit{kv-v2} is enabled at path \textit{cmsweb}. We used HashiCorp's \textit{Agent Injector} in the CMSWEB cluster as their \textit{CSI Provider} doesn’t provide secret templating, which we require. The working of the agent injector is shown in Fig. \ref{fig:vault_k8s_working}. A Vault Agent container that renders Vault secrets to a shared memory volume is added to the pod specs by the \textit{Vault Sidecar Agent Injector} using the sidecar pattern. Containers in the pod can then consume Vault secrets without being Vault-aware by rendering secrets to a shared volume. A Kubernetes mutating webhook controller serves as the injector. Only if there are annotations in the request, the controller intercepts pod events and applies modifications to the pod Fig \ref{fig:vault_k8s_working}.

\begin{figure}[!t]
\includegraphics[scale=0.4]{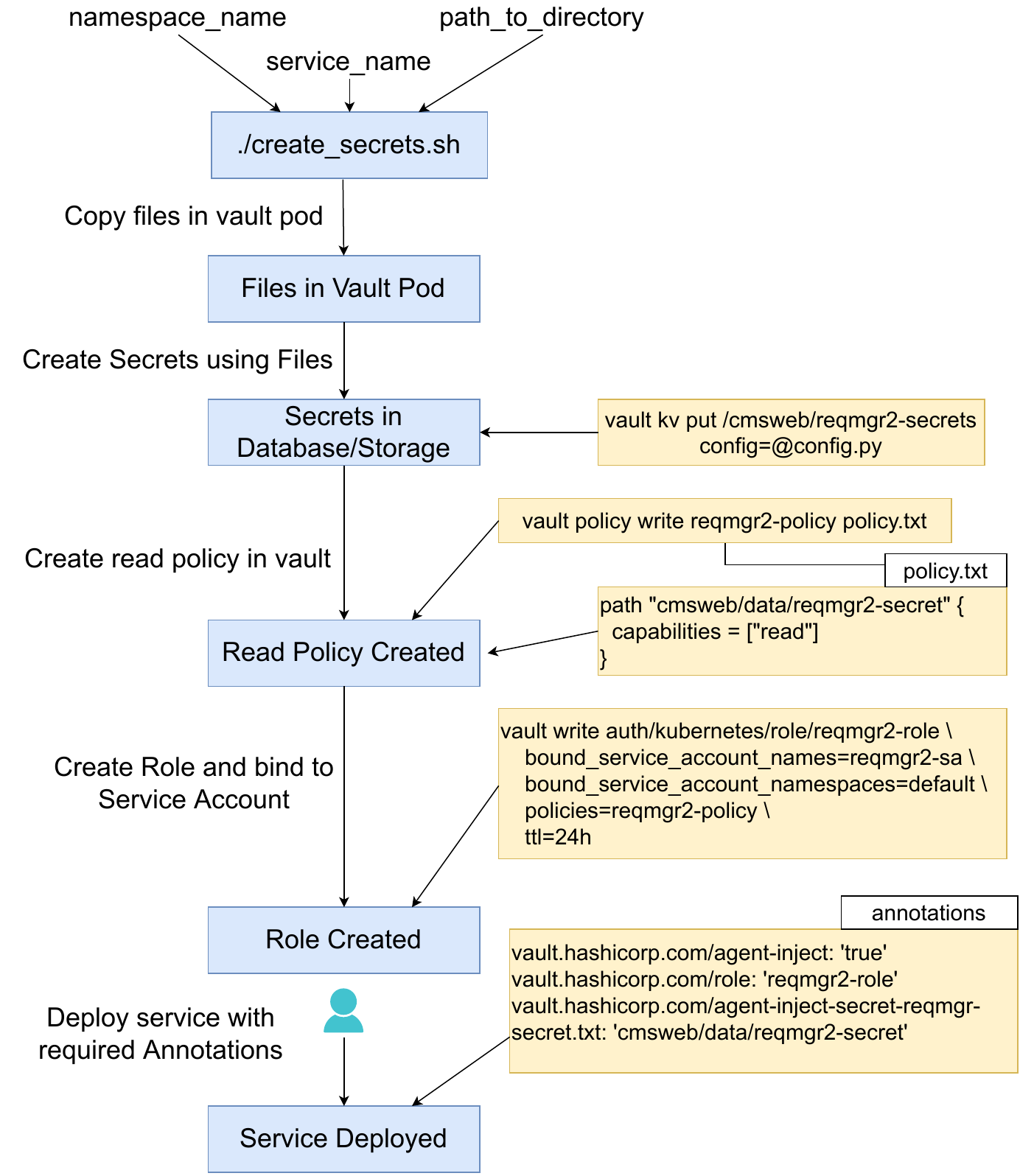}
\centering
\caption{Workflow of Vault Script}
\label{fig:secrets_script}
\end{figure}

Creating secrets in the Vault gets more complicated when secrets need to be created from files. To create secrets in CMSWEB, we have created a script. The worflow of the script is shown in \ref{fig:secrets_script}. The script requires 3 arguments: the \textit{namespace}, \textit{service}, and \textit{path} to the directory that contains files. It copies files into the vault pod so it's accessible by the Vault CLI. Then it creates a secret using the service name and appending it with "-secrets". The secret is required to be a key-value pair, so names of files are used as keys and content as values. The script creates a policy that allows reading the secret. Finally, the script creates a role to bind policy with the service account. After running the script, the user is required to add annotations to the service pods for the Vault Agent Injector.

Adding Vault to the CMSWEB cluster brought on multiple benefits to the cluster. Authorization of the service before accessing a secret is the more prominent one, though it also decouples secrets from services that make services scalable. It also makes secret management centralized and easy. The audit functionality of Vault helps to find possible anomalies. Vault also allows security policies like "a secret can only be accessed once a day" and others so we can configure policies according to our requirements in the future.

\section{Conclusion}
\label{chap:Conclusion}

In this paper, we presented work related to the addition of new security features to the CMSWEB Kubernetes cluster. These include enforcement of network policies, the deployment of OPA Gatekeeper policies with custom helm charts and the deployment of Vault to manage secrets.

Network policies are in-house guards against code vulnerabilities. By default, Kubernetes allows every pod to communicate with every other pod in the cluster, so if one pod is compromised, all pods have to be considered as being compromised. With the network policies in place, even if a malicious actor compromises one service, they won't be able to compromise another service. This also prevents the propagation of any sort of virus or ransomware.

OPA Gatekeeper policies further ensures that nobody can trick the system as they implement fine-grained control over different activities that is not possible otherwise. Even if somebody gets access to the cluster somehow, it won't be easy to install any payload or do any malicious activity as long as OPA Gatekeeper is in place. This also prevents unintentional and unfavorable activity in the cluster. The OPA Gatekeeper enables in-depth control over the environment, policies, and activities.

Vault encrypts the secrets and tightly controls access to the secrets so a user, machine, or service has to authorize itself first before accessing the secrets. This way, the secret is accessible to the only service that is authorized to view it. Also, the secrets and sensitive information are decoupled from the cluster, making the system scalable and allows the same secret to be accessible to more than one service, simplifying the management of the secrets.

\bibliography{biblio}
\end{document}